\documentclass[a4paper]{jpconf}
\usepackage{amsmath}
\usepackage{amssymb}
\usepackage{graphicx}
\usepackage[square,sort&compress,numbers]{natbib}
\usepackage{chngpage}
\newcommand*{\mpl}{M_{\rm{Pl}}}
\newcommand*{\lag}{\mathcal{L}}

\begin{document}

\title{Phase space analysis of the $F (X) - V (\phi)$ scalar field Lagrangian and scaling solutions in flat cosmology}

\author{Josue De-Santiago$^{1,2,3}$ and Jorge L. Cervantes-Cota$^2$}

\address{$^1$Universidad Nacional Aut\'onoma de M\'exico, 04510, D. F., M\'exico}
\address{$^2$Depto. de F\'{\i}sica, Instituto Nacional de Investigaciones Nucleares, M\'{e}xico}
\address{$^3$Institute of Cosmology $\&$ Gravitation, University of Portsmouth, PO1 3FX, United Kingdom}
\ead{josue@ciencias.unam.mx, jorge.cervantes@inin.gob.mx}

\begin{abstract}
  We review a system of autonomous differential equations developed
  in our previous work \cite{DeSantiago:2012nk} describing a flat cosmology filled
  with a barotropic fluid and a scalar field with a modified kinetic term of the
  form $\lag=F(X)-V(\phi)$. We analyze the critical points and summarize 
  the conditions to obtain scaling solutions. We consider a
  set of transformations and show that they leave invariant the equations
  of motion for the systems in which the scaling solution is obtained, allowing to
  reduce the number of degrees of freedom.
\end{abstract}

\section{Introduction}

In cosmology, the scalar fields play an important role due to its broad phenomenology, which
can be used in order to describe different phenomena. For that reason they  
have been used to model phenomena like
inflation \cite{Liddle:2000cg}, dark energy \cite{Copeland:2006wr,delaMacorra:1999ff}, dark matter
\cite{Magana:2012ph}, bounce scenarios \cite{Peter:2008qz},
and unification models \cite{Bertacca:2010ct}.

The canonical Lagrangian for a scalar field is given by
 $\lag = X- V(\phi)$,
where the first term $X=-\frac{1}{2}\partial_\mu \phi \partial^\mu \phi$ is the
kinetic term, and the second is the potential that can have different functional
forms depending on the model.
In the last years, a generalization
has been a studied in which the Lagrangian is a general function of $X$ and $\phi$.
These Lagrangians
were used initially in the study of inflation \cite{ArmendarizPicon:1999rj}
and dark energy \cite{Chiba:1999ka}, and in a previous work
we used this type of Lagrangian to unify inflation, dark matter, and dark
energy \cite{DeSantiago:2011qb}.

One particular form of this kind of Lagrangian is the sum-separable, in which
\begin{equation}
 \lag=F(X)-V(\phi), 
  \label{lag}
\end{equation}
 and there is a clear separation between the
kinetic term $F(X)$ and the potential term $V(\phi)$. This class of Lagrangians has 
the advantage of being more easily studied than the more general case while still
conserving some of the rich phenomenology that is not present in
the canonical case.
In a previous work we studied the dynamical system of this class of Lagrangians
in a flat Friedmann-Lemaitre-Robertson-Walker (FLRW) cosmology filled with the field
and an additional barotropic fluid \cite{DeSantiago:2012nk}.
This analysis allowed us to find the critical points of the system and its stability, from
which we drew conclusions about the general behaviour and evolution of the system. 

In the present work we will review this analysis and we will add an extra analysis for
the case in which the system has non-trivial critical points. We will show that in
that case there is a symmetry that allows the degrees of freedom of the system to
reduce. This behaviour is similar to what happens in the canonical
Lagrangian with exponential potential. This symmetry allows the existence of scaling
solutions even when the Lagrangian is not of the form $\lag=Xg(Xe^{\lambda\phi})$, 
which in the literature \cite{Piazza:2004df} is considered to be the most general
form of Lagrangians with this behaviour.

\section{Autonomous system}

Along this work we will consider a flat FLRW cosmology
filled with a scalar field with Lagrangian of the form (\ref{lag}) and a
barotropic fluid with equation of state $P_m=(\gamma_m-1)\rho_m$ where $\gamma_m$ is constant.
In this system, if no interaction between the field and the fluid is considered, the
equations of motion correspond to the two independent Einstein equations, also called Friedmann
equations and the two continuity equations as follows:
\begin{eqnarray}\label{fried1}
   && H^2 = \frac{1}{3\mpl^2} [2XF_X -F + V + \rho_m] \,,
   \\ && \label{fried2}
   \frac{dH}{dt} = - \frac{1}{2\mpl^2} [2XF_X  + \gamma_m \rho_m] \,,
   \\ && \frac{d\rho_m}{dt} = - 3H\gamma_m \rho_m \,,
   \\ \label{cont} &&
   \frac{d}{dt} (2XF_X - F+V) = -6 H XF_X \,,
\end{eqnarray}
where the subscript $F_X$ means differentiation of $F$ with respect to $X$.
Here only three of the equations are independent from each other, as
the system has only three degrees of freedom. We can choose an initial
$\phi$, $\dot{\phi}$ and $\rho_m$,
which in turn will define the initial Hubble parameter with the equation
(\ref{fried1}) and the initial kinetic term $X$ with the relation
\begin{equation}
  X = \frac{1}{2} \dot \phi ^2\,.
  \label{Xdef}
\end{equation}

In order to obtain an autonomous system of differential equations we define as
independent variable the dimensionless $dN=d\log a$,
and then define the new dynamical variables
\begin{eqnarray}
  x &=& \sqrt{\frac{\rho_k}{\rho_c}} \,, \\
  y &=& \sqrt{\frac{V}{\rho_c}} \,, \\
  \sigma &=& -\frac{\mpl}{\sqrt{3\rho_k}}\frac{d\log V}{dt} \,,
\end{eqnarray}
where $\rho_k=2XF_X-F$ is the density associated to the kinetic part of the Lagrangian,
and $\rho_c=3\mpl^2 H^2$ is the critical density for a FLRW cosmology.
The evolution equation (\ref{fried1}) is written in terms of the 
new defined dynamical variables as
$x^2+y^2+\Omega_m = 1$, where $\Omega_m=\rho_m/\rho_c$ is positive,
which implies that both variables $x$ and $y$ are bounded between 0 and 1.

Using the previous relation and the evolution equations (\ref{fried1}-\ref{cont}),
we find the
autonomous differential equations in terms of the dynamical variables as 
\begin{eqnarray}
   \label {dxdN} \frac{dx}{dN} &=&
     \frac{3}{2} \sigma y^2 +
     \frac{3}{2} x \left[ \gamma_k (x^2 - 1) + \gamma_m (1 - x^2 - y^2 ) \right] \,,
   \\ \label{dydN} 
   \frac{dy}{dN} &=&
     \frac{3}{2}y \left[ - \sigma x +
     \gamma_m (1 - y^2 ) + 
     x^2 (\gamma_k - \gamma_m) \right] \, ,
   \\  \label{DSDN}
   \frac{d\sigma}{dN} &=&  -3\sigma^2 x (\Gamma -1) +
     \frac{3\sigma (2\Xi \gamma_k + \gamma_k - 2)}{2\gamma_k(2\Xi+1)}
     \left( \gamma_k - \frac{\sigma y^2}{x} \right),
\end{eqnarray}
where we required the introduction of the auxiliary variables
\begin{eqnarray}
  \gamma_k &=&  \frac{\rho_k + P_k}{\rho_k} = 
  \frac{2XF_X}{2XF_X-F} \,,
  \\
  \Xi &=&  \frac{X F_{XX}}{F_X} \,,
  \\
  \Gamma &=& \frac{VV_{\phi\phi}}{V_\phi^2} \,.
\end{eqnarray}
These auxiliary variables depend on derivatives of the potential
and kinetic terms in the
Lagrangian and in general are not constant, which implies that in order
to solve the system of equations (\ref{dxdN}-\ref{DSDN}) it is necessary
to consider evolution equations for the extra three variables. These
equations in general will depend on higher derivatives of the Lagrangian
terms which in turn will have evolution equations.

In order to cut the succession of equations it is possible to demand the Lagrangian
to satisfy that both $\gamma_k$ and $\Gamma$ be constant, which implies that
\begin{equation} \label{kinpow}
  F(X)= AX^\eta \,,
\end{equation}
and
\begin{eqnarray}
  V(\phi)&=& B(\phi-\phi_0)^n \, \,, \, {\rm or} \label{potpow}
  \\
  V(\phi)&=& V_0 e^{-\lambda\phi} \,, \label{potexp}
\end{eqnarray}
where $A$, $B$, $V_0$, $\eta$, $n$, and $\lambda$ are constants. In this case
$\gamma_k=2\eta/(2\eta-1)$, $\Xi=\eta-1$, and $\Gamma$ is one if  
the potential is exponential or $(n-1)/n$ if it is a power-law.
In this particular Lagrangian the system of equations
(\ref{dxdN}-\ref{DSDN}) is closed and we can study its solutions without concern
about other equations, as we will do in the next section.

\section{Critical points}\label{criticalpoints}

The critical points of the system are obtained when the equations
(\ref{dxdN}-\ref{DSDN}) 
are equal to zero.
Besides analyzing the conditions for the existence of the critical points we
determine their stability. Equating only the equations (\ref{dxdN})
and (\ref{dydN}) to zero, we obtain the critical points (a), (b), ($\alpha$), ($\beta$)
and ($\gamma$) summarized in table \ref{tab:1}, where the first two are present 
in the canonical case and have been analized in previous works like
\cite{Copeland:1997et, Gumjudpai:2005ry} and the last three are new.
Another two critical points are found for a particular case of the Lagrangian,
when it can be written either as a canonical Lagrangian with exponential potential
(\ref{potexp}) or as the sum of two power-law terms (\ref{kinpow}, \ref{potpow}),
where their exponents satisfy
\begin{equation}
  \eta = \frac{n}{2+n}\,.
  \label{sym}
\end{equation}
These critical points are named (c) and (d), and written in table \ref{tab:1} too. As the values for
$x$ and $y$ in these points are dependent on $\sigma$, for them to be fixed it is also
necessary for the evolution equation (\ref{DSDN}) to be zero, in \cite{DeSantiago:2012nk}
we showed that this is the case when (\ref{sym}) is satisfied or the Lagrangian is a
canonical with exponential potential. In the next section we will add more elements to this
analysis and we will show that in fact $\sigma$ is a function of $x$ and $y$ for
these Lagrangians.

\begin{table}
  \caption{Stability and existence of the critical points. The 
  points labeled with Latin letters can be reduced in the canonical case to the ones
  already studied in the literature, the points with Greek letters are new. The
  points (c) and (d) exist only when (\ref{sym}) holds or when the Lagrangian is
  canonical with exponential potential.}
  \label{tab:1}
\begin{adjustwidth}{-1em}{}
\begin{tabular}{ccccccc}
    &$x$ & $y$ & Existence & Stability & $\Omega_\phi$ & $\gamma_\phi$ \\
    \hline \hline
    (a) & 0 & 0 & Always & Unstable node for $\gamma_k<\gamma_m$ &0 & - \\
    & &  & &Saddle point for $\gamma_m<\gamma_k$ & & \\

    \hline 
    ($\alpha$) &0 & 1 & $\sigma=0$ & Saddle point for $\gamma_k<0$ & 1 & $0$ \\
    & &  & &Stable node for $\gamma_k>0$ & & \\

    \hline
    (b) &1 & 0 & Always & Unstable node for
    $\displaystyle \gamma_k>\left\{ \gamma_m, \sigma  \right\}$& 1&$\gamma_k$ \\
    && & &Stable node for
    $\displaystyle\gamma_k<\left\{ \gamma_m, \sigma \right\}$ & & \\
    && & &Otherwise saddle point & & \\

    \hline
    ($\beta$) &0 & Arbitrary & $\sigma=0$ & Stable line for $\omega_k>-1$&$y^2$ & $0$ \\
    & & & $\gamma_m=0$ & Unstable otherwise & & \\

    \hline
    ($\gamma$) &Arbitrary&0&$\gamma_k=\gamma_m$&Stable line for $x\sigma>\gamma_k$ & $x^2$&$\omega_m$ \\
    &&&&Unstable otherwise&&\\
    
    \hline
    (c) &$\displaystyle \frac{\sigma}{\gamma_k}$ &
    $\displaystyle \sqrt{1 - \frac{\sigma^2}{\gamma_k^2}}$&
    $\displaystyle \sigma^2>\gamma_k^2$  &
    Saddle point for $\displaystyle \sigma^2>\gamma_k \gamma_m$
    & 1 & $\displaystyle \frac{\sigma^2}{\gamma_k}$\\
    && &$\sigma \gamma_k>0$ & Otherwise stable node
    & & \\

    \hline
    (d) &$\displaystyle \frac{\gamma_m}{\sigma}$&
    $\displaystyle \frac{\sqrt{\gamma_m (\gamma_k-\gamma_m)}}{\sigma}$&
    $\sigma>\sqrt{\gamma_k \gamma_m}$ &
    Stable node for
    $\displaystyle \frac{\sigma^2(9\gamma_m-\gamma_k)}{8\gamma_k \gamma_m^2}<1$& 
    $\displaystyle \frac{\gamma_k \gamma_m}{\sigma^2}$& $\gamma_m$\\
    &&&$\gamma_m<\gamma_k$&Stable spiral otherwise&&\\

    \hline
  \end{tabular}
\end{adjustwidth}
\end{table}

\section{Symmetry}

As we showed in the previous section, when the Lagrangian of the scalar field is the
sum of two power-law terms and the relation (\ref{sym}) holds,
two extra critical points
are present in the system. In this section we will see that for this system exists
a symmetry in the equations of motion (\ref{fried1}-\ref{cont}). This symmetry permits
the number of degrees of freedom to be reduced to only two. 

The equations of motion (\ref{fried1}-\ref{cont}) written for a Lagrangian
with power-law terms (\ref{kinpow}, \ref{potpow}) and
with $dN$ as independent variable instead of $dt$ are given by
\begin{eqnarray}
  \label{fried1_aux}  &&
  H^2 = \frac{1}{3\mpl^2} [ (2\eta-1)A X^\eta + B\phi^n + \rho_m] \,,
   \\ \label{fried2_aux} &&
   H \frac{dH}{dN} = - \frac{1}{2\mpl^2} [2\eta A X^\eta  + \gamma_m \rho_m] \,,
   \\ &&
   \frac{d\rho_m}{dN} = - 3 \gamma_m \rho_m \,,
   \\ \label{cont_aux} &&
   \frac{d}{dN} ( (2\eta-1)AX^\eta + B \phi^n) = -6 \eta A X^\eta \,.
\end{eqnarray}
If we consider $\phi$, $X$, and $\rho_m$ as the independent variables we can apply the
transformation
\begin{eqnarray}
  \phi &\rightarrow & \xi^{2\eta} \phi \,, \nonumber \\
  X &\rightarrow& \xi^{2n} X \,, \nonumber \\
  \rho_m &\rightarrow & \xi^{2n\eta} \rho_m \,,
  \label{transf}
\end{eqnarray}
where if we leave $N$ invariant, from the 
definition of $X$
\begin{equation}
  X=\frac{1}{2} \left( H \frac{d\phi}{dN} \right)^2,
  \label{Xdef2}
\end{equation}
the Hubble parameter transforms as $H \rightarrow \xi^{n-2\eta}H$. This
transformation will leave invariant the equations of motion
(\ref{fried1_aux}-\ref{cont_aux}) as long as $n$ and $\eta$ satisfy the relation
(\ref{sym}), which determines the transformation of the Hubble parameter to be
$H \rightarrow \xi^{n \eta}H$.

The fact that the equations of motion are invariant under the transformations definded in (\ref{transf}) for arbitrary $\xi$ 
means that this set of transformations is a symmetry of the system, and allows to
reduce the number of degrees of freedom to two. The only thing needed is to define a set of variables
that are invariant under the transformation, but $x$, $y$ and $\sigma$ 
are already invariant. Moreover, there are only two invariant
degrees of freedom, which implies that one of them is dependent of the other. In
fact we can see that the relation
\begin{equation}
  \sigma =-\mpl n B^{1/n} \sqrt{\frac{2}{3}}(A(2\eta-1))^{-1/2\eta} \left( \frac{x}{y} \right)^{2/n}\,,
  \label{bla}
\end{equation}
holds when (\ref{sym}) is satisfied. If we define
\begin{equation}
  s \equiv - \mpl n B^{1/n} \sqrt{\frac{2}{3}}(A(2\eta-1))^{-1/2\eta} \,,
\end{equation}
the dynamical system can be written from (\ref{dxdN}, \ref{dydN}) as
\begin{eqnarray}
  \label {dxdN2} \frac{dx}{dN} &=&
     \frac{3}{2} x \left[ sy \left( \frac{y}{x}\right)^{2/\gamma_k} +
     \gamma_k (x^2 - 1) + \gamma_m (1 - x^2 - y^2 ) \right] \,,
     \\
  \label{dydN2}  \frac{dy}{dN} &=&
     - \frac{3}{2} s x^2 \left( \frac{y}{x} \right)^{2/\gamma_k} +   
     \frac{3}{2} y \left[ \gamma_m (1 - y^2 ) + 
     x^2 (\gamma_k - \gamma_m) \right] \, ,
\end{eqnarray}
corresponding to only two equations for two variables.
Here we used the fact that $\gamma_k = 2n/(n-2)$ when the relation (\ref{sym})
is satisfied.

\subsection{Exponential potential}

The behaviour stated in the this section is similar to what happens for the
canonical Lagrangian with exponential potential (\ref{potexp}), and studied in Ref. \cite{Holden:1999hm}.
In that case a similar procedure to the one used in this section can tell us that
the equations of motion are invariant under the transformations
\begin{align}
    \phi \rightarrow \phi + \xi \,,  &&
    X \rightarrow e^{-\lambda \xi} X \,, &&
    \rho_m \rightarrow e^{-\lambda \xi} \rho_m \,. \label{transcan}
\end{align}
The fact that the system is invariant under the transformations above,
allows it to be described by only two variables as in the original work
of this type of cosmological analysis \cite{Copeland:1997et}.
In that case, unlike the relation (\ref{bla}), $\sigma$
is a constant given by $\sqrt{2/3}\mpl \lambda$.

\section{Phase space}

The critical points found in section \ref{criticalpoints}
are still valid for the 2-dimensional system
(\ref{dxdN2}, \ref{dydN2}), as can be seen by equating the right-hand side of both equations
to zero. The critical points (c) and (d) depend on $\sigma$ that is function of $x$ and $y$
so we should be able to drop de dependence on $\sigma$ using
the relation (\ref{bla}). For (c) it cannot be made analytically while for
(d) we obtain:
\begin{align}
  x_d = \frac{\sqrt{\gamma_m(\gamma_k-\gamma_m)}}{s} 
    \left( \frac{\gamma_m}{\gamma_k- \gamma_m} \right)^{1/\gamma_k} \,,
  &&
  y_d = \frac{\gamma_k - \gamma_m}{s} 
    \left( \frac{\gamma_m}{\gamma_k-\gamma_m} \right)^{1/\gamma_k} \,.
  \label{xdclosed}
\end{align}

\begin{figure}
   \centering
   \includegraphics[width=8cm,keepaspectratio=true]{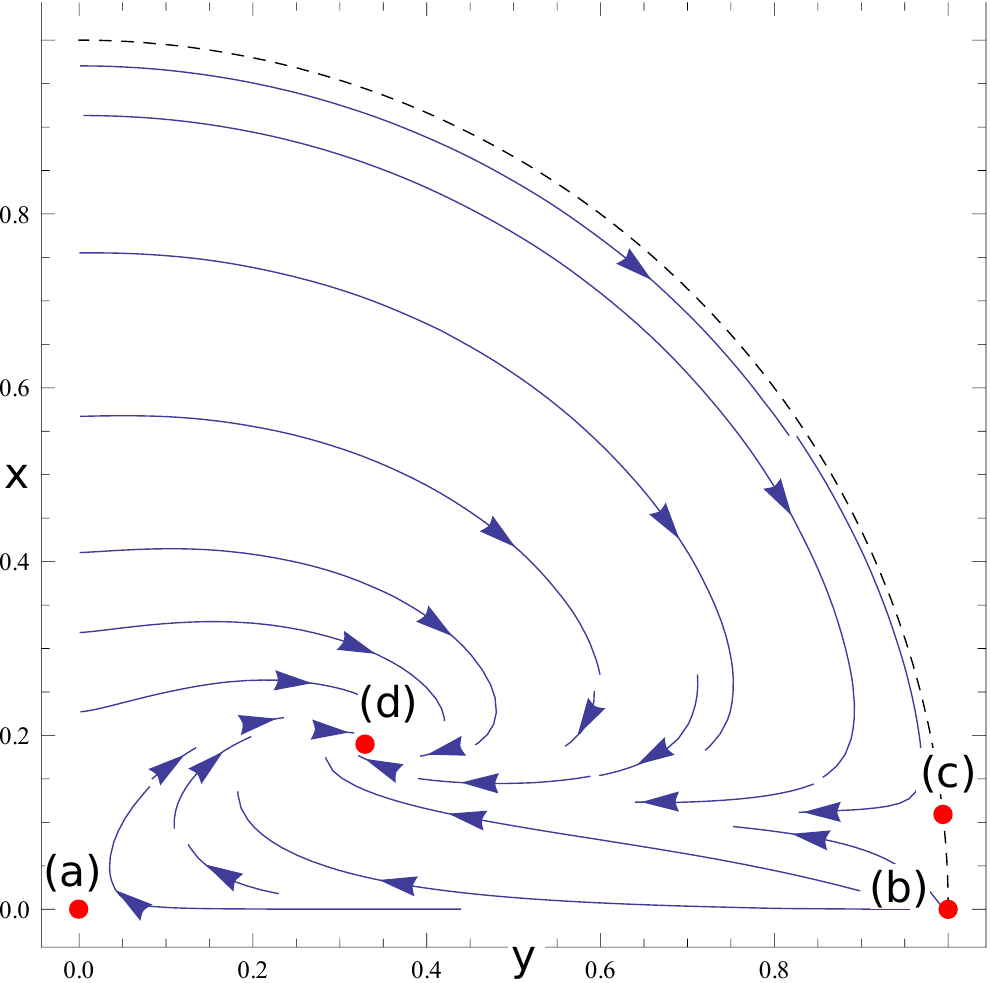}
   \caption{Phase space for the dynamical system corresponding to $\lag=AX^2 + B \phi^{-4}$.
   The critical points (a), (b), (c)  and (d) can be observed.}
   \label{fig1}
\end{figure}

The behaviour of the critical points stated in the table \ref{tab:1} can be seen more graphically
in the phase space plot in figure \ref{fig1}.
There we see the phase space of a system with
Lagrangian
\begin{equation}
  \lag = AX^2 + B \phi^{-4} \,,
\end{equation}
plus a dust type barotropic fluid with $\gamma_m=1$, where we chose $s=4$ meaning that $AB=4\mpl^4/27$.
In the figure we can observe the critical points:
\begin{itemize}
  \item (a) The (0,0) point that in this case is a saddle point.
  \item (b) The (1,0) point that in this case is an unstable node.
  \item (c) Corresponding to a saddle point, with domination of field density over
    the barotropic fluid.
  \item (d) The point $(3^{1/4}/4,3^{-1/4}/4)$ corresponding to a stable spiral, and whose value can be
    computed from expression (\ref{xdclosed}) for the values of this particular example.
\end{itemize}
The other critical points are not present in this example.

\section{Conclusions}

In this work we made an analysis of the dynamical system associated to a flat FLRW cosmology
containing a barotropic fluid and a scalar field with Lagrangian of the form $\lag= F(X)-V(\phi)$.
We observed that this system can be described by three variables and its equations of motion, from
those we defined $x$, $y$ and $\sigma$ and obtained their evolution equations. For Lagrangians
with constant values for $\gamma_k$ and $\Gamma$, the three autonomous differential equations (\ref{dxdN} - \ref{DSDN}) are 
enough to describe the system. For more general Lagrangians however, it is necessary
to obtain dynamical equations for the variables $\gamma_k$, $\Gamma$, and $\Xi$,
that in turn will include new variables in terms of higher derivatives of the
Lagrangian,
that will require extra equations of motion. In order to have only the minimal
three equation system, in our work \cite{DeSantiago:2012nk}
we used power-law kinetic terms  that
produce $\gamma_k$ constant, and power-law or
exponential potentials that correspond to a constant value of $\Gamma$.

We obtained the critical points of the system, and studied their
existence and stability conditions, which are summarized in table \ref{tab:1}. The critical
points (a), (b), (c), (d) can be reduced for the canonical Lagrangian
to those studied in previous works \cite{Wands:2008tv, Gumjudpai:2005ry},
meanwhile ($\alpha$), ($\beta$) and ($\gamma$)
are new.

In our previous work \cite{DeSantiago:2012nk} we showed that for the existence of the
critical points (c) and (d), the
Lagrangian needs to be either canonical with exponential potential or 
the sum of two power-law terms that satisfy the relation (\ref{sym}). 
In the present work we showed that when this relation holds,
the equations of motion (\ref{fried1_aux} - \ref{cont_aux})
are invariant under the set of transformations defined by (\ref{transf}). This symmetry
allows to reduce the number of degrees of freedom of the system to two. We showed that $x$ and $y$
are invariant under the same transformation (\ref{transf}) and hence are suitable to describe
the system. Therefore, it was possible to obtain $\sigma$ as function of $x$ and $y$ in equation
(\ref{bla}). Using this equation we found the two dimensional autonomous system (\ref{dxdN2}, \ref{dydN2}).

We showed that for the two dimensional system
the critical points are the same as the ones obtained in table \ref{tab:1}, but with
$\sigma$ as a function of $x$ and $y$. The critical point (c) 
can be obtained then as a function of the auxiliary variables $\gamma_k$ and
$\gamma_m$ only, eliminating the dependence on $\sigma$, however it is not possible to do it
analytically. For the critical point (d) the analytical expression can be obtained
as we have done in (\ref{xdclosed}).

The symmetry transformations of (\ref{transf}) are similar to those of (\ref{transcan}), corresponding to canonical
exponential Lagrangians and studied in \cite{Holden:1999hm}. Both transformations allow the systems to be treated with a two
dimensional set of dynamical equations. For the canonical exponential Lagrangian the equations get reduced to
(\ref{dxdN}, \ref{dydN}) with a constant value of $\sigma$, while for the Lagrangian
whose terms satisfy (\ref{sym}), they get reduced to equations (\ref{dxdN2}, \ref{dydN2}).

\ack
JDS is supported by ININ Grant.

\providecommand{\newblock}{}

\end{document}